\documentclass[twocolumn]{aastex63}
\usepackage{gensymb}
\usepackage{natbib}

\usepackage[utf8]{inputenc}
\usepackage[maxfloats=1000]{morefloats}
\usepackage{comment}
\shorttitle{Sicilian et al.}
\shortauthors{Sicilian et al.}

\turnoffedit

\begin{document}
\title{Constraining Sterile Neutrino Dark Matter in the Milky Way Halo with Swift-XRT}

\email{d.sicilian@miami.edu}

\author{Dominic Sicilian}
 \affiliation{Department of Physics, University of Miami, Coral Gables, Florida 33146, USA}

\author{Dannell Lopez}
\affiliation{Department of Physics, University of Miami, Coral Gables, Florida 33146, USA}
\affiliation{School for Advanced Studies -- Wolfson Campus, Miami, Florida 33132, USA}

\author{Massimo Moscetti}
\affiliation{Department of Physics, University of Miami, Coral Gables, Florida 33146, USA}
\affiliation{Vanderbilt University, Nashville, Tennessee 37240, USA}
\affiliation{Palmer Trinity School, Miami, Florida 33157, USA}

\author{Esra Bulbul}
\affiliation{Max Planck Institute for Extraterrestrial Physics, Garching bei München, Bayern, Germany}

\author{Nico Cappelluti}
\affiliation{Department of Physics, University of Miami, Coral Gables, Florida 33146, USA}

\begin{abstract}

We present a search for sterile neutrino dark matter decay signals in the Milky Way's dark matter halo by considering the entirety of the Swift-XRT data archive. After filtering the raw archive, we analyze a $\sim$77 Ms data set containing the full field of view, as well as a $\sim$41 Ms data set with point-sources excised using the Swift-XRT Point Source catalog. We report non-detections of emission lines across the 3--6 keV continuum in both data sets, including at and around 3.5 keV. The point-sources excised data set is found to have higher sensitivity to faint dark matter decay signals due to its freedom from point-source contamination and is thus used to set constraints. Non-detections across the total data set's continuum are used to constrain the sterile neutrino dark matter parameter space, marginally strengthening existing X-ray constraints. Non-detections at $\sim$3.5 keV in data subsets grouped by angular distance from the galactic center are used to constrain the 3.5 keV line's galactic intensity profile, providing the strongest constraints to date across $\sim$1/4 of the galaxy.

\end{abstract}

\section{Introduction}

The sterile neutrino is a popular dark matter (DM) candidate, as it is a natural extension to the Standard Model and can be tested using current X-ray observatories (see, e.g., \citealt{dodwid 1994}; \citealt{aba 2001a}; \citealt{aba 2001b}; \citealt{laine 2008}; \citealt{boyarsky 2019}). This testability is due to the particle's decay mechanism, dictated by the mixing angle sin$^2(2\theta)$, which releases a photon with $E = m_{\nu_s}/2$ (where $m_{\nu_s}$ is the sterile neutrino mass; \citealt{pal 1982}) and is therefore within X-ray bands for $m_{\nu_s} \sim 10$ keV.

Sterile neutrino DM garnered heavy attention when \cite{esra 2014} observed an unidentified emission line at $\sim$3.5 keV in various galaxy clusters, found to be consistent with $\sim$7 keV sterile neutrino decay. Afterwards, the 3.5 keV line was repeatedly detected in several DM-dominated objects such as clusters, M31, and the galactic center (GC) using multiple X-ray telescopes (\citealt{boyarsky 2014}; \citealt{wik 2014}; \citealt{boyarsky 2015}; \citealt{urban 2015}; \citealt{neronov 2016}; \citealt{franse 2016}; \citealt{nico 2018}; \citealt{hofmann 2019}). Other X-ray works did not detect unidentified emission at 3.5 keV or any other energy, setting constraints on the sin$^2(2\theta)$ vs. $m_{\nu_s}$ parameter space (\citealt{suzaku 2015}; \citealt{ng 2015}; \citealt{bulbul 2016}; \citealt{oleg 2016}; \citealt{hitomi 2017}; \citealt{perez 2017}; \citealt{dessert 2020}; \citealt{bhargava 2020}; \citealt{sicilian 2020}; \citealt{silich 2021}; \citealt{foster 2021}; \citealt{roach 2022}). In particular, the keV-mass parameter space has been most strongly constrained by works that used large-scale blank-sky data sets from major X-ray observatories including Chandra \citep{sicilian 2020}, \edit1{XMM-Newton} \citep{dessert 2020, foster 2021}, and NuSTAR (\citealt{neronov 2016}; \citealt{roach 2020}; \citealt{roach 2022}), which make use of the ubiquitous, DM-dominated Milky Way Halo (MWH).

Here, we perform a similar MWH analysis using Swift-XRT, an instrument chosen for \edit1{three reasons}. First, its archive is vast and yet-untapped for MWH sterile neutrino constraints, and thus we sought to continue to push current technology to its limits by mining the archive of another major telescope. Secondly, \edit1{it is important to test data from all X-ray observatories to test the instrumental origins of putative features such as the 3.5 keV line. And third,} despite its relatively low grasp, Swift's low-Earth orbit affords it a substantially lower and more stable non-X-ray particle background (NXB) than Chandra or \edit1{XMM-Newton}, potentially giving it improved sensitivity to a sterile neutrino DM signal in the $m_{\nu_s} \sim $ 5--10 keV range, where X-ray constraints have typically been dominated by Chandra and \edit1{XMM-Newton}.

In this work, we consider the entirety of the Swift-XRT data archive up to 2021. This unfiltered data set contains $\sim$188 Ms of exposure time and, due to the short average exposure of Swift observations, consists of $\sim$2 orders of magnitude more observations than \cite{sicilian 2020}'s $\sim$51 Ms Chandra data set, thus giving nearly all-sky coverage. We hence aim to use Swift's low NXB, the large data set, and its wide spatial distribution to constrain both the sterile neutrino DM parameter space and the galactic intensity profile of the 3.5 keV line. Our methodology closely follows that of \cite{sicilian 2020}, which is thoroughly described in that work and summarized here.

\section{Data Selection and Reduction}

To minimize contamination from extended sources, we excluded observations pointed through the galactic plane ($| b |\leq10^\mathrm{o}$) or within $2^\mathrm{o}$ of major known extended sources, including the Crab Nebula and any object in either the ROSAT galaxy cluster catalog (ROSGALCLUS; \citealt{rosat clusters}) or Chandra supernova remnant catalog\footnote{\url{https://hea-www.harvard.edu/ChandraSNR/}}. After applying these filters, the data set contained $\sim$77 Ms of exposure time.

Upon applying standard Swift-XRT data reduction procedures\footnote{\url{https://swift.gsfc.nasa.gov/analysis/xrt_swguide_v1_2.pdf}}, we replicated \cite{moretti 2009}'s methodology to extract both field of view (FOV) and NXB spectra from individual observations. In particular, the NXB was extracted from regions of the CCD not exposed during observations, while the FOV spectrum was taken from \cite{moretti 2009}'s conservatively-defined central $\sim$0.054 deg$^2$ region to avoid contamination from out-of-time-event calibration.

To produce a data set with minimal point-source contamination, for any observation considered by the Swift-XRT Point Source catalog (2SXPS; \citealt{2SXPS}), we excised all documented sources in that observation \edit1{according to the catalog-supplied point-spread function (PSF) at each source position}. These point-sources excised (PSE) observations form a $\sim$41 Ms data subset. \edit1{This process was not as efficient as \cite{sicilian 2020}'s, since Chandra's PSF is superior to Swift's, but the much-lower continuum in the PSE compared to FOV shows we removed considerable contamination regardless.} Excising point-sources inadvertently removes DM along the line-of-sight (LOS), and the lack of coverage offered by 2SXPS reduced our total exposure time by nearly 50\%, so therefore, like \cite{sicilian 2020}, we analyzed both the $\sim$41 Ms PSE and the full $\sim$77 Ms FOV data sets.

We then binned the total PSE and FOV data sets by angular distance from the galactic center ($\theta_{\mathrm{GC}}$) to study the 3.5 keV line's intensity profile as it compares to Navarro-Frenk-White (NFW; \citealt{nfw}) predictions and prior measurements. Bins were determined by emulating \cite{sicilian 2020}'s estimation of the minimum Chandra exposure time needed to detect the 3.5 keV line as a function of $\theta_{\mathrm{GC}}$, which we modified to accommodate Swift. \cite{sicilian 2020}'s line profile constraints, while strong, were difficult to interpret near the GC, where NFW slope is steep, as that work was grouped into only 4 bins, with the innermost bin spanning 10--74$^\mathrm{o}$. Therefore, although Swift's lower NXB relative to Chandra's already relaxes bin exposure thresholds, we further amplified this by binning the data sets based on their exposure time after removing the galactic plane but before removing any other observations with extended sources, rather than their fully-filtered exposures. As a result, we may have sacrificed some strength in our profile constraints, but we were nonetheless able to greatly improve $\theta_{\mathrm{GC}}$ resolution, with the PSE and FOV data sets yielding 9 and 13 bins, respectively. 

Spectra were stacked using \texttt{mathpha} (from \texttt{FTOOLS}) to produce master FOV, PSE, and NXB spectra, as well as to produce the corresponding sets of $\theta_{\mathrm{GC}}$-binned spectra. Exposure-weighted response files were combined using \texttt{FTOOLS} \texttt{addrmf} and \texttt{addarf}. NXB spectra were renormalized to match each respective data set's total counts in channels $\geq$725, based on NXB-to-total spectral counts ratios computed in \cite{moretti 2009}.

\section{Spectral Analysis}\label{analysis}

\begin{figure}[h!] 
\includegraphics[width=9.cm]{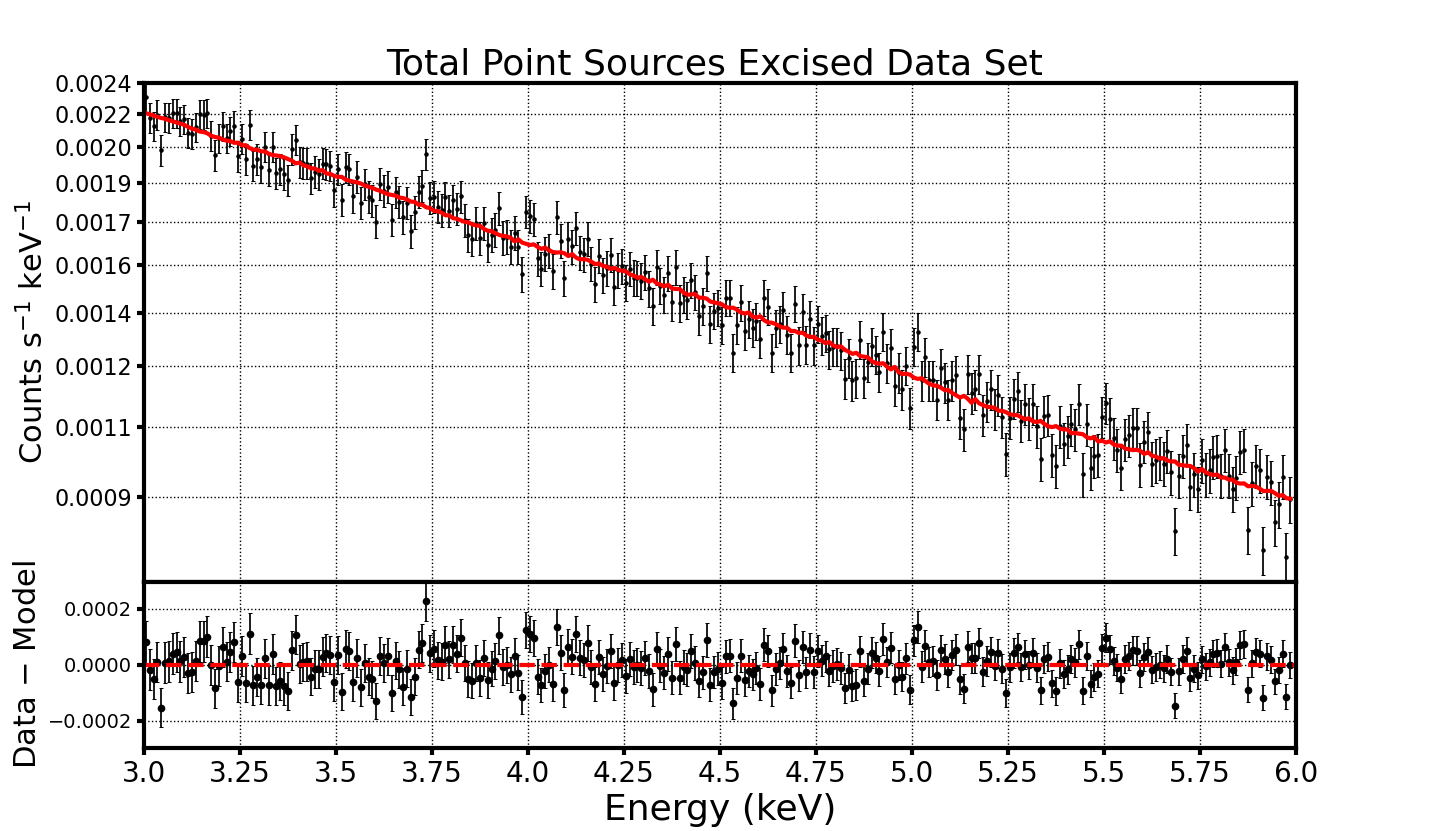}
\includegraphics[width=9.cm]{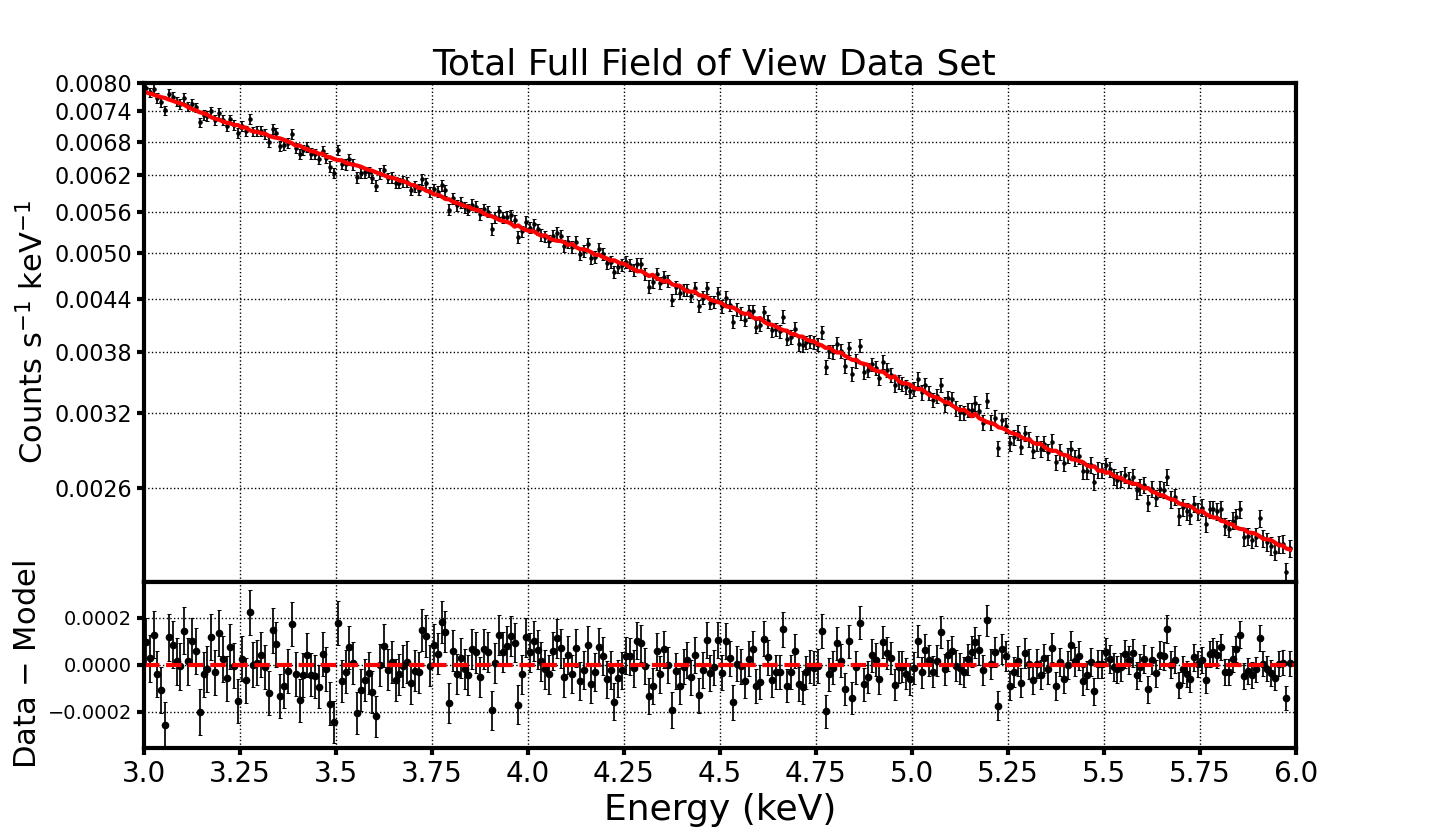}
\caption{Spectra and best-fit models for the total PSE (\textbf{top}) and FOV (\textbf{bottom}) data sets.}
\label{fig:modelplots}
\end{figure}

Spectra were modeled using \texttt{PyXspec} (\texttt{XSPEC} 12.12.0) on the interval 3.0--6.0 keV, chosen to include 3.5 keV while avoiding known emission features \citep{moretti 2009} and soft galactic diffuse emission, as well as giving more coverage than \cite{sicilian 2020}. As shown in \cite{moretti 2009}, the total blank-sky continuum in this band consists of $\sim$34\% cosmic X-ray background (CXB), $\sim$16\% off-axis stray-light (SL) due to Swift's lack of Chandra/XMM-like baffles, and $\sim$51\% NXB. Spectra for the total PSE and FOV data sets are plotted in Figure \ref{fig:modelplots} along with best-fit models produced through the procedure described below.

The astrophysical continuum was modeled using a broken power law folded through both the ARF and RMF which, like \cite{moretti 2009}'s CXB power law, is unabsorbed due to the low $N_H$ across fields considered \citep{dickey 1990} and its resulting lack of absorption in our model's energy band. The NXB model was a broken power law folded only through the RMF, as the NXB arises strictly from particles that do not pass through Swift-XRT's mirrors (\citealt{pagani 2007}; \edit1{\citealt{esra 2020}}). Visual inspection of the spectra revealed possible emission features at $\sim$3.75 and $\sim$5.88 keV, and our band also includes energies of known features at 3.3 and 3.68 keV modeled in \cite{sicilian 2020} whose origins--whether astrophysical or instrumental--are unclear. However, we can conclude from preliminary \texttt{CSTAT} \citep{cash 1979} optimization fits that, for both the FOV and PSE data sets, there are no significant features at these energies in either the astrophysical or NXB spectra, as we found $\Delta$BIC-obtained significances approaching zero for those features, while no $\Delta\chi^2$-computed significance exceeded $\sim$2$\sigma$. These results were consistent regardless of whether line width was fixed or left free. Therefore, we do not include any such features in our final models.

Best-fit values for the reported models (Table \ref{tab:parameters}) were obtained using Goodman-Weare MCMC fitting (with \texttt{CSTAT} as the fit statistic) to generate posterior parameter distributions. The resulting best-fit models for the total FOV and PSE spectra have reduced $\chi^2$ ($\chi_{\nu}^2$) of $\sim$1.1, and all $\theta_{\mathrm{GC}}$-binned models have $\chi_{\nu}^2 \sim$ 1.1--1.2, indicating the models are good fits.

\begin{table}
{\renewcommand{\arraystretch}{1.2}
\begin{tabular}{c || c | c  }

\hline
  & FOV & PSE \\
\hline\hline
$\Gamma_{1}$ &1.84$_{-0.04}^{+0.04}$ & 2.70$_{-0.20}^{+0.19}$ \\
$E_b$ (keV)&3.98$_{-0.07}^{+0.07}$ & 3.96$_{-0.10}^{+0.11}$ \\ 
$\Gamma_{2}$ &1.21$_{-0.05}^{+0.05}$ & -0.10$_{-0.24}^{+0.28}$ \\ 
$\mathrm{log}K$ &-3.30$_{-0.02}^{+0.02}$ & -3.85$_{-0.11}^{+0.10}$ \\ 
\hline
$\Gamma_{1, \mathrm{NXB}}$ &0.66$_{-0.09}^{+0.09}$ & 0.64$_{-0.08}^{+0.07}$\\
$E_{b,\mathrm{NXB}}$ (keV) &5.31$_{-0.12}^{+0.09}$ & 5.23$_{-0.14}^{+0.12}$ \\ 
$\Gamma_{2, \mathrm{NXB}}$ &-0.62$_{-0.21}^{+0.22}$ & -0.38$_{-0.23}^{+0.19}$\\
$\mathrm{log}K_{\mathrm{NXB}}$ &-2.36$_{-0.06}^{+0.05}$ & -2.45$_{-0.05}^{+0.05}$\\
\hline\hline
$\chi_{\nu}^2$ & 1.148 & 1.055\\
\hline

\end{tabular}}
\caption{Model parameter best-fit and 1$\sigma$ error values from MCMC posteriors for the total PSE and FOV data sets. $\chi_{\nu}^2$ is also reported to indicate goodness-of-fit. The top panel shows astrophysical parameters while the bottom shows NXB values. $\Gamma$ represents photon indices, $E_b$ is the break energy, and $K$ indicates broken power law normalizations.}
\label{tab:parameters}
\end{table}

\subsection{The 3.5 keV Line}

For sterile neutrino constraints, we first focused on the known 3.5 keV feature following \cite{sicilian 2020}'s procedure. We added an emission line with $E$ free to vary from 3.4--3.6 keV, \edit1{with width fixed at zero and thus dictated} by the instrument response, then performed another MCMC fit, the posteriors of which revealed that $E$ is generally poorly-constrained, with no convergence in the FOV line $E$ posterior and mild convergence in PSE line $E$. Both had the same best-fit $E = 3.48$ keV and similar best-fit line fluxes log$F_{line} \sim$ -8. We isolated a feature with frozen $E = 3.48$ keV and performed an MCMC fit (with walkers initialized for log$F_{line}$ between -8 and -7 to match the $E$-free best-fit $F_{line}$'s order of magnitude, then left free to vary across all values). Using the resulting best-fit $F_{line}$, we assessed the significance of the feature using both $\Delta\chi^2$ and $\Delta$BIC, finding it to approach zero for the total FOV and PSE models. The same results were found when repeating the procedure with \cite{sicilian 2020}'s frozen $E = 3.51$ keV. We therefore conclude that this is a non-detection of the 3.5 keV line.

We applied this method to all $\theta_{\mathrm{GC}}$ bins in both the FOV and PSE data sets, adopting $E = 3.51$ keV for our frozen line $E$ as in \cite{sicilian 2020}, due to its use in that work, consistency with past detection energies, and proximity to the center of the allowed interval when we left line $E$ free to vary due to the sporadic, inconsistent shapes and nature of convergences of the resulting $E$ posterior distributions. The 3.51 keV feature yielded near-zero $\Delta$BIC-computed significances for all $\theta_{\mathrm{GC}}$ bins in both the PSE and FOV data sets and near-zero $\Delta\chi^2$ significances in all PSE bins, as well as in nearly all FOV bins, with no FOV bin exceeding $\sim$1$\sigma$. These results thus further confirm our conclusion that we did not detect a 3.5 keV line.

\subsection{The Continuum}

\begin{figure}[h!] 
\includegraphics[width=9.cm]{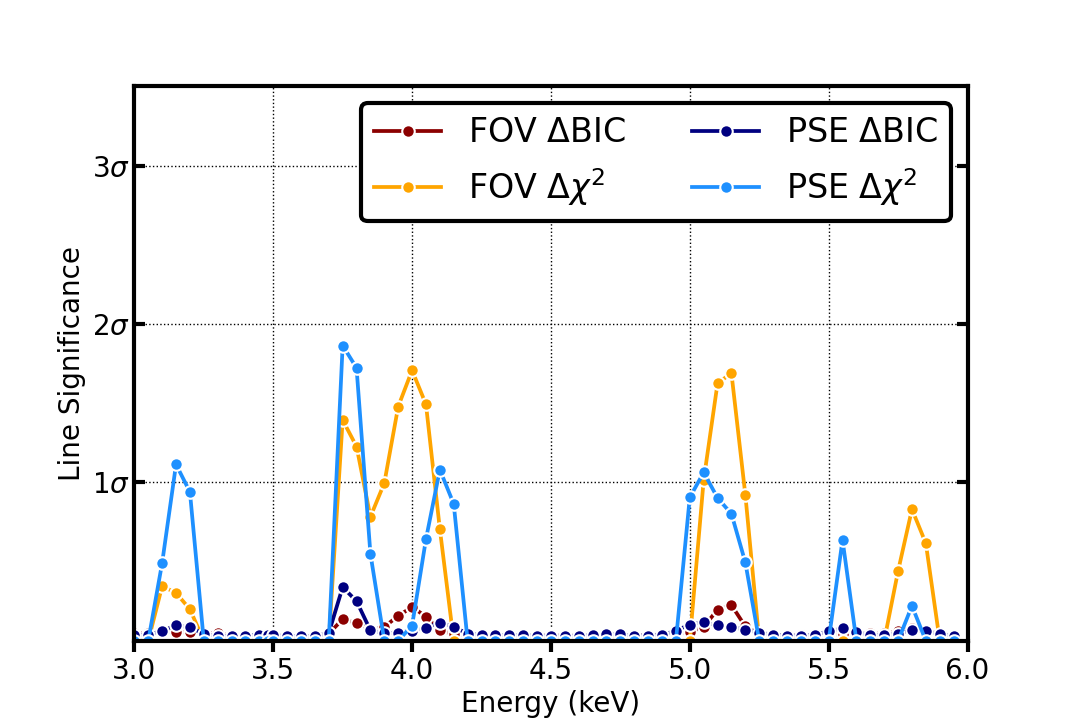}
\caption{Data points indicate line significances when fitted at the discrete energies analyzed across the continuum. Connecting lines are used to enhance visualization.}
\label{fig:significance}
\end{figure}

To search for faint emission lines at other model energies, we performed the same process above across the remainder of the continuum. From 3--6 keV, we increased the frozen line $E$ in steps of 50 eV (chosen to be smaller than Swift's energy resolution and hence protect against skipping over a genuine feature) and used the MCMC-computed best-fit $F_{line}$ to evaluate significance. In Figure \ref{fig:significance}, we have plotted line significance computed using $\Delta$BIC and $\Delta\chi^2$ at all energies in both the FOV and PSE data sets. As shown, all $\Delta$BIC-obtained significances were $\ll$1$\sigma$. For nearly all energies in both models, $\Delta\chi^2$ significance was also $\ll$1$\sigma$, and in all remaining cases was $<$2$\sigma$. Therefore, we conclude that we did not detect any emission features consistent with sterile neutrino DM.

The 3$\sigma$ upper-limit on the line flux at each energy is computed as the 0.997 quantile from that energy's $F_{line}$ MCMC posterior distribution, representing the 99.7\% confidence interval above $F_{line} = 0$. We computed these limits for both the total FOV and PSE data sets, providing two sets of flux upper-limits at all energies considered across the continuum, including 3.48 and 3.51 keV. This method was also used to set upper-limits on the $\theta_{\mathrm{GC}}$ 3.5 keV line flux for the 9 PSE bins and 13 FOV bins.

We compared our continuum constraints to a series of simulations to ensure the robustness of our upper-limit calculations, particularly to verify that the limits are not misleadingly low, which could conceivably occur if the MCMC walkers reach low $F_{line}$ values early in the fit and remain localized throughout. The simulation procedure follows that of \cite{sicilian 2020}, which repeatedly simulated an emission feature at each energy to compute the maximum $F_{line}$ value that would have yielded a $\Delta\chi^2$ significance $<$3$\sigma$ and hence managed to avoid detection. We computed these constraints for both the FOV and PSE spectra, then compared them to those obtained using MCMC. In both cases, the upper-limits generated by the simulations were more aggressive than the MCMC constraints by $\sim$24\%, decisively confirming the validity of our results. Thus, we can now use our line flux upper-limits to constrain the 3.5 keV line's $\theta_{\mathrm{GC}}$ profile and compute sterile neutrino parameter space constraints.

\section{Results and Discussion}

Despite the FOV's exposure time advantage and PSE's loss of DM along the LOS due to point-source removal, the PSE's line flux upper-limits for energies across the continuum are $\sim$17\% lower on average than the FOV's flux limits, and its $\theta_{\mathrm{GC}}$ bins yield $\sim$29\% more stringent profile constraints than those of the FOV. This is due to the higher sensitivity afforded to the PSE data by eliminating baryonic contamination from point sources and therefore minimizing continuum flux. For a faint DM signal, even when accounting for PSE's LOS losses, we estimate the PSE's signal-to-noise ratio (S/N) to be $\sim$22\% greater than that of the FOV data. It follows that the PSE data set should yield similarly more stringent constraints, which is highly consistent with our results. Therefore, like \cite{sicilian 2020}, we have chosen to employ the PSE data set to determine DM constraints.

\subsection{3.5 keV Line Intensity Profile Constraints}

\begin{figure}[h!] 
\includegraphics[width=9.cm]{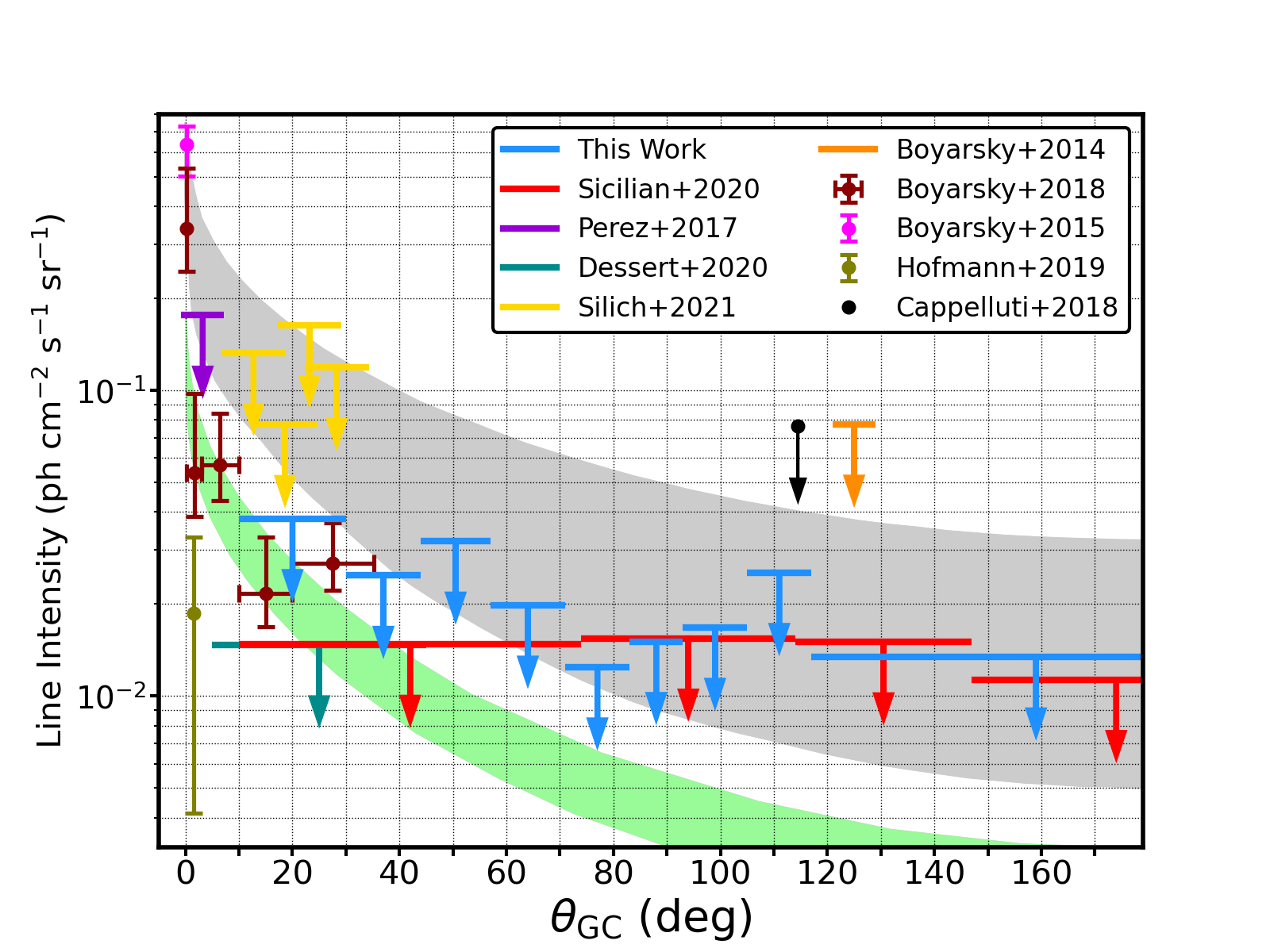}
\caption{The Swift line intensity upper-limit profile plotted alongside existing constraints, prior detections, and two example NFW-predicted profiles for reference. The gray profile was computed by \cite{nico 2018} \edit1{(using \citealt{nesti 2013}'s parameters)} while the green profile was computed by \cite{boyarsky 2018} \edit1{(using \citealt{mcmillan 2017}'s parameters)}. Shaded regions for both represent 2$\sigma$ uncertainty ranges. Adapted from \cite{sicilian 2020}.}
\label{fig:NFW}
\end{figure}

Using our 3.5 keV line flux upper-limits in the $\theta_{\mathrm{GC}}$ bins, we have computed and plotted the intensity upper-limit profile. Existing constraints and prior detections are shown in Figure \ref{fig:NFW}, with the exception of \cite{neronov 2016}'s NuSTAR detections, as their decaying DM origins were ruled out by \cite{sicilian 2020}'s profile constraints. Two example NFW-predicted profiles are shown for reference, computed by \cite{nico 2018} and \cite{boyarsky 2018} based on different sets of NFW parameters and normalized by \cite{boyarsky 2015}'s GC detection. As shown in Figure \ref{fig:NFW}, our constraints on the intensity profile are the strongest to date across $\sim$1/4 of the galaxy, producing the most stringent limits in the regions 71--93$^\mathrm{o}$ and 117--147$^\mathrm{o}$, with \cite{sicilian 2020}'s generally remaining the lowest everywhere else $\geq$10$^\mathrm{o}$.

As also illustrated by Figure \ref{fig:NFW}, the higher $\theta_{\mathrm{GC}}$ resolution relative to \cite{sicilian 2020} allows a more useful interpretation near the GC than the stronger \cite{sicilian 2020} upper-limit. Similarly, the \cite{dessert 2020} XMM MWH constraint (shown here with \citealt{sicilian 2020}'s adjustment based on \citealt{abazajian 2020}, \citealt{boyarsky 2020}, and \citealt{dessert 2020a}) is nearly identical to \cite{sicilian 2020}'s in that region, but has $\sim$1/2 the Swift constraints' $\theta_{\mathrm{GC}}$ resolution. The Swift constraints improve upon \cite{silich 2021}'s HaloSat limits with our 10--30$^\mathrm{o}$ bin's combination of strength and resolution relative to those of \cite{silich 2021}'s results.

However, by the same token, the \cite{dessert 2020} constraint \edit1{may be more informative than ours in that region and appears inconsistent with the decaying DM origin of \cite{boyarsky 2018}'s detections between 10--35$^\mathrm{o}$, though there has been extensive discussion \citep{abazajian 2020, boyarsky 2020, dessert 2020a} challenging that limit due to concerns over \cite{dessert 2020}'s exclusion of 3.3 and 3.68 keV features in the final model, which heavily impacted that work's constraints. Notably, we also excluded those features as they were not detected (Section \ref{analysis}), but ensured the robustness of our results by recomputing $\sim$3.5 keV constraints when including 3.3 and 3.68 keV features in the total PSE model, finding that the line flux upper-limit differed by just $\sim$2\% at 3.51 keV and by less than 1\%, on average, across the 3--4 keV band. Moreover, interestingly, the \cite{boyarsky 2018} detections are entirely consistent with our profile constraints}.

\subsection{Parameter Space Constraints}

Due to the direct proportionality between sin$^2(2\theta)$ and the sterile neutrino decay rate for a given $m_{\nu_s}$, namely
\begin{equation}\label{decay_rate}
\Gamma_\gamma(m_{\nu_s},\theta) = 1.38 \times 10^{-29} \ \mathrm{s}^{-1} \frac{\mathrm{sin}^2(2\theta)}{10^{-7}} \bigg({\frac{m_{\nu_s}}{1 \  \mathrm{keV}}}\bigg)^5,
\end{equation}
where $\Gamma_\gamma(m_{\nu_s},\theta)$ is the decay rate \citep{pal 1982}, and the predicted flux of decaying DM emission ($F_{\mathrm{DM}}$) from a given mass of DM in the FOV ($M_{\mathrm{DM,FOV}}$) a distance $D$ from the detector, given by
\begin{equation}\label{DM_flux}
F_{\mathrm{DM}} = \frac{\Gamma_\gamma}{4\pi m_{\nu_s}}\frac{M_{\mathrm{DM,FOV}}}{D^2}
\end{equation}
\citep{neronov 2016b}, determining sin$^2(2\theta)$ constraints from line flux upper-limits is dependent on the distribution of DM observed within the data set. The constraints, then, must be obtained by integrating the DM density profile over the LOS and FOV and are therefore model-dependent on the profile function and its parameter values. As in \cite{sicilian 2020}, when computing the continuum constraints, we performed the calculation twice--first using fairly aggressive NFW parameters based on the empirical \cite{nesti 2013} values (scale radius $R_s \sim 16$ kpc; local DM density $\sim 0.67$ GeV cm$^{-3}$) and again using conservative parameters detailed in \cite{abazajian 2020} ($R_s = 26$ kpc; $\rho_\odot \sim 0.28$ GeV cm$^{-3}$).

\begin{figure}
\includegraphics[width=9.cm]{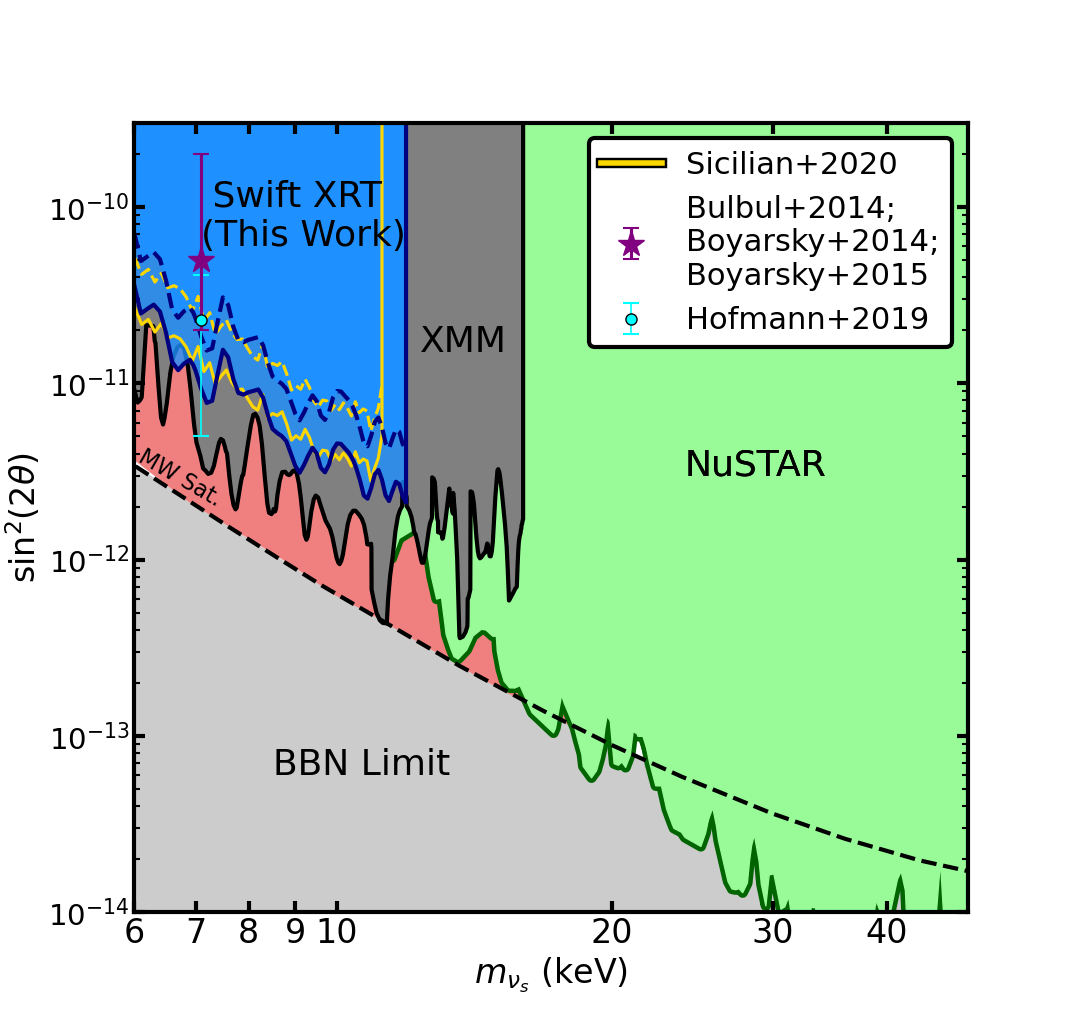}
\caption{Constraints on the sin$^2$(2$\theta$) vs. $m_{\nu_s}$ parameter space. The red ``MW Sat." region represents \cite{dekker 2021b}'s study of small-scale structure, the dark gray ``XMM" region indicates \cite{foster 2021}, and the green ``NuSTAR" region represents \cite{neronov 2016}, \cite{perez 2017}, \cite{ng 2019}, \cite{roach 2020}, and \cite{roach 2022}. Dashed gold represents the conservative \cite{sicilian 2020} Chandra constraint, while solid gold is the more aggressive constraint, with the gap between them representing NFW uncertainty. As described in the text, our Swift constraints are plotted in a similar way, with the uncertainty region appearing translucent while the region above the dashed conservative limit is solid. Purple and cyan data points show notable $\sim$3.5 keV line detections. Adapted from \cite{ng 2019}, \cite{roach 2020}, and \cite{sicilian 2020}.}
\label{fig:roach}
\end{figure}

The resulting Swift constraints on sin$^2(2\theta)$ are plotted in Figure \ref{fig:roach} along with other existing limits on the parameter space. Like \cite{sicilian 2020}, we expect the true constraint to lie somewhere between the bounds generated by the different NFW parameter choices, so we have plotted the conservative curve as a dashed line, separated from the more aggressive curve by a translucent uncertainty region. As mentioned when computing continuum line flux upper-limits, unlike in \cite{sicilian 2020} these constraints were designed to include coverage for the 3.5 keV line, eliminating the need for a separate point on the plot.

As Figure \ref{fig:roach} shows, despite the Swift exposure being $\sim$20\% less than \cite{sicilian 2020}'s $\sim$51 Ms of Chandra exposure, the constraints are highly consistent and in fact our Swift constraints are slightly stronger overall, with the conservative limits $\sim$4\% more stringent than the conservative Chandra constraints and the aggressive limits $\sim$1\% more stringent than the aggressive Chandra constraints. Notably, at $m_{\nu_s} \sim 7$ keV, our conservative constraints are just outside the collective 1$\sigma$ error range of the \cite{esra 2014}, \cite{boyarsky 2014}, and \cite{boyarsky 2015} detections, providing evidence against the 3.5 keV line's \edit1{decaying} DM interpretation. However, even our aggressive constraints remain consistent with \cite{hofmann 2019}'s detection, which means our results do not entirely eliminate the 3.5 keV line's \edit1{decaying} DM interpretation, though \cite{foster 2021}'s results (which are stronger than ours at $m_{\nu_s} \sim 7$ keV) are outside even the \cite{hofmann 2019} 1$\sigma$ error range. \edit1{The overall DM origin of the 3.5 keV line also remains feasible in frameworks besides the sterile neutrino that may be observed in clusters but not galaxies (e.g., \citealt{cline 2014a}; \citealt{cline 2014b}; \citealt{cline 2014c}; \citealt{fink 2016}).} While the \cite{foster 2021}'s XMM MWH constraints are also more stringent overall than ours, towards the aggressive end of our limits we were able to match the \edit1{XMM-Newton} constraints at several $m_{\nu_s}$ values and improve upon \edit1{XMM-Newton}'s around $\sim$6.7 and $\sim$12 keV.

Therefore, the edge of the overall X-ray limits is now formed by constraints from Swift in this work, XMM in \cite{foster 2021}, and NuSTAR in a combination of \cite{neronov 2016}, \cite{perez 2017}, \cite{ng 2019}, \cite{roach 2020}, and \cite{roach 2022}. Together, these X-ray constraints have eliminated most of the parameter space that remains above the big bang nucleosynthesis (BBN) lower limits, which are theoretical limits based on lepton asymmetry in the early Universe (see \citealt{dolgov 2002}; \citealt{serpico 2005}; \citealt{laine 2008}; \citealt{boyarsky 2009}; \citealt{ven 2016}; \citealt{roach 2020}). According to \cite{dekker 2021b}, this small remainder of the parameter has been ruled out nearly entirely up to $m_{\nu_s} \sim 20$ keV based on that work's study of small-scale structure formation around the MWH, though those results are dependent on systematics such as MWH mass, NFW parameters, and the extended Press-Schechter formalism (EPS; \citealt{eps1}; \citealt{eps2}).

\subsection{Closing Thoughts}

We have found no evidence of sterile neutrino DM in the Swift-XRT MWH observations, and have used our non-detections to constrain the 3.5 keV line's galactic intensity profile, as well as the sterile neutrino DM sin$^2(2\theta)$ vs. $m_{\nu_s}$ parameter space. Our 3.5 keV line profile upper-limits are the strongest to date across $\sim$1/4 of the galaxy and our parameter space limits marginally improve upon existing X-ray constraints. Swift has now joined Chandra, \edit1{XMM-Newton}, and NuSTAR on the list of major X-ray observatories whose vast archives have been harnessed in the search for decaying DM emission lines in the MWH. Since the constraints obtained from these observatories are fairly consistent, it appears we have collectively pushed existing technology and data to their limits, suggesting recent or future X-ray missions such as XRISM will be necessary to resolve lingering uncertainty on this topic. This sentiment is evidenced by \cite{dekker 2021a}'s simulations indicating eROSITA will achieve the required sensitivity to do so. Until then, the door will remain nearly closed, yet undeniably ajar, on sterile neutrino dark matter.

\acknowledgements

DS kindly acknowledges NASA Swift Grant 80NSSC21K 1410 and the University of Miami for providing funding during the completion of this work. The authors thank the University of Miami Young Scholars Program for facilitating the contributions of future scientists to works such as this.

\software{Astropy (\citealt{astropy 2013}; \citealt{astropy 2018}), corner \citep{corner}, HEASoft 6.29 \citep{heasoft}, Matplotlib \citep{matplotlib}, NumPy \citep{numpy}, pandas (\citealt{pandas}; \citealt{pandas2}), PyXspec \citep{pyxspec}, Scikit-learn \citep{pedregosa 2012}, XSPEC 12.12.0 \citep{xspec}}

\end{document}